\documentclass[
aps,%
12pt,%
final,%
notitlepage,%
oneside,%
onecolumn,%
nobibnotes,%
nofootinbib,%
superscriptaddress,%
noshowpacs,%
centertags]%
{revtex4}

\usepackage[english]{babel}
\usepackage[utf8]{inputenc}
\usepackage[T1]{fontenc}
\usepackage{amsmath}
\usepackage{amssymb}
\usepackage{graphicx}
\usepackage{hyperref}
\usepackage{color}
\usepackage{bm}
\usepackage{braket}

\begin{document}

\title{Few-particle lepton bound states in variational approach}

\author{A.~V.~Eskin}
\author{A.~P.~Martynenko}
\author{F.~A.~Martynenko}
\email{f.a.martynenko@gmail.com}
\author{D.~K.~Pometko}
\email{pometko-darja@yandex.ru}

\affiliation{Samara University, 443086, Moskovskoye shosse 34, Samara, Russia}

%\date{\today}

\begin{abstract}
The energy levels of the ground states of the three-particle and four-particle bound states of leptons in quantum electrodynamics are calculated. For the calculation, the variational method with Gaussian basis functions is used. The hyperfine structure of the spectrum is taken into account due to the pairwise spin-spin interaction of particles.
\end{abstract}


\maketitle

\section{Introduction}

Few-lepton systems (positronium, positronium ion, positronium molecule, muonium molecule, etc.) consist 
of charged leptons that interact mainly through electromagnetic forces, and their fundamental theory is 
quantum electrodynamics (QED) \cite{p1,p2}. Due to the absence of a structure such as that of the proton, 
leptonic systems have occupied an important place in elementary particle physics. The study of their energy 
levels and decay widths has allowed for testing the gauge theory of particle interactions with high accuracy 
and determining the values of fundamental physical constants. For many decades, mainly two-particle leptonic 
states (positronium and muonium) were investigated, but later three-particle (positronium ion) \cite{p4} and 
four-particle (positronium molecule, muonium molecule) systems began to be studied \cite{p5,p5a}. These exotic 
systems are of interest not only to theoretical physicists but also to researchers in atomic and molecular 
physics, allowing the study of the complex mechanism of formation of large groups of particles rather than 
splitting them into small parts \cite{p5b,p5c,p5d}.

For systems with a small number of leptons, it is of interest to study not only bound states but also 
resonance states. Studies of resonant four-particle states with leptons have been conducted to a significantly 
lesser extent to date. Including muons, along with electrons, in this area of research opens up new prospects, 
although this is associated with certain difficulties. It is known that calculating resonant states requires 
a number of specialized methods and significantly greater computational resources than calculating 
bound states \cite{apm2025,p6}. In recent years, the physics of muonic atoms and molecules has developed 
quite successfully. Studies of the energy levels of light two-particle and three-particle muonic atoms 
of hydrogen and helium have been completed. The plans of a number collaborations, among others, 
include new tasks involving new muonic atoms. 
The use of high-intensity and polarized muon beams makes it possible to study the dynamics of muons and 
muonium at the atomic level \cite{muon}.
The use of new muon beams in various research centers also opens up new possibilities in the study of 
three-lepton and four-lepton systems with muons, which can be created by introducing a muon beam 
into an electron gas \cite{p11a}.

When studying bound states of particles in different quantum theories of electromagnetic and strong 
interactions, common patterns in their description often arise, which allow us to draw analogies 
between seemingly completely different systems.
Few-lepton bound states are formed as a result of the Coulomb interactions at short distances between particles. 
Such bound states are somewhat similar to the bound states of four quarks (tetraquarks), which were discovered 
in 2003 by the Belle Collaboration and have been intensively studied over the past two decades \cite{p10,p10aa}. 
In heavy tetraquarks, the quarks themselves have low kinetic energy and are located at short distances. 
This leads to the fact that the color-electric interaction between quarks becomes dominant, just as 
in a tetralepton system the Coulomb interaction is dominant. Therefore, a bound system of four leptons 
in quantum electrodynamics and a bound system of four heavy quarks in quantum chromodynamics 
can be considered similarly in an attempt to find some general patterns and gain a deeper understanding 
of the nature of tetraquark states.
However, we also note that the main difference between such systems is that in the case of bound states, 
an attractive potential is usually chosen between individual heavy quarks, whereas in Coulomb systems, 
both attractive and repulsive forces act between charged particles.
The aim of our work is to develop the variational method used in \cite{p8,p9} for the case of four particles 
with Coulomb interactions. We study a number of lepton bound states and calculate the binding energies, 
including hyperfine interaction effects.

\section{General formalism}

Let us consider the description of the Coulomb bound states of four particles within 
the variational method in coordinate representation.
Let ${\bf r}_1$, ${\bf r}_2$, ${\bf r}_3$, ${\bf r}_4$ be the radius vectors of the particles 
in the initial reference frame.
We now move on to the Jacobi coordinates $\boldsymbol\rho$, $\boldsymbol\lambda$, $\boldsymbol\sigma$, 
which are related to the initial coordinates in the center-of-mass system as follows:
\begin{equation}
\mathbf{r}_1 = -\frac{m_4}{m_{1234}}\boldsymbol{\sigma} - \frac{m_3}{m_{123}}\boldsymbol{\lambda} - \frac{m_2}{m_{12}}\boldsymbol{\rho}, 
\mathbf{r}_2 = -\frac{m_4}{m_{1234}}\boldsymbol{\sigma} - \frac{m_3}{m_{123}}\boldsymbol{\lambda} + \frac{m_1}{m_{12}}\boldsymbol{\rho},
\end{equation}
\begin{displaymath}
\mathbf{r}_3 = -\frac{m_4}{m_{1234}}\boldsymbol{\sigma} + \frac{m_{12}}{m_{123}}\boldsymbol{\lambda}, \nonumber \\
\mathbf{r}_4 = \frac{m_{123}}{m_{1234}}\boldsymbol{\sigma},
\end{displaymath}
where $m_i$ is the mass of the $i$-th particle, $m_{ij}=m_i+m_j$, $m_{1234}=\sum_{i=1}^4 m_i$.

\begin{figure}[htbp]
\centering
\includegraphics[scale=1.2]{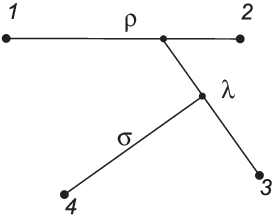}
\caption{Jacobi coordinates for a system of four particles.}
\label{fig1}
\end{figure}

This choice of the Jacobi coordinates is shown in Fig.~\ref{fig1}. The coordinate $\rho$ denotes the relative distance between 
the pair of particles 1 and 2. The coordinate $\lambda$ denotes the relative distance between particle 3 and the center 
of mass of particles 1 and 2. The coordinate $\sigma$ denotes the relative distance between particle 4 and the center 
of mass of particles 1, 2, 3.

In the nonrelativistic approximation, the Hamiltonian of the four-particle system contains the kinetic energy operator 
and pairwise Coulomb interaction potentials:
\begin{equation}
\hat{H} = -\frac{1}{2m_1}\Delta_1 - \frac{1}{2m_2}\Delta_2 - \frac{1}{2m_3}\Delta_3 - \frac{1}{2m_4}\Delta_4 + 
\sum_{i,j=1,i>j}^4\frac{e_i e_j}{r_{ij}},
\label{f2a}
\end{equation}
where $e_i$ are the electric charges of the particles, $r_{ij}$ are the distances between the charges $i$ and $j$. Transforming 
into the Jacobi coordinates, the kinetic energy operator can be represented in the center-of-mass system 
as a sum of three terms, each containing the Laplace operator for the corresponding variables:
\begin{equation}
\hat{T} = \frac{\mathbf{p}^2_\rho}{2\mu_1} + \frac{\mathbf{p}^2_\lambda}{2\mu_2} + \frac{\mathbf{p}^2_\sigma}{2\mu_3},
\label{f2}
\end{equation}
where the reduced masses of the particles $\mu_1$, $\mu_2$, $\mu_3$ are determined by the following expressions:
\begin{equation}
\mu_1 = \frac{m_1 m_2}{m_{12}}, \quad \mu_2 = \frac{m_3 m_{12}}{m_{123}}, \quad \mu_3 = \frac{m_{4} m_{123}}{m_{1234}}.
\label{f3}
\end{equation}

For finding the system's wave function by the variational method, the exponential or Gaussian form is usually used 
\cite{p6,p7,p7a}. We choose the variational functions for the ground state of the four-particle system 
in the Gaussian form, as in previous works \cite{p8,p9,p10,p10a,p10b}:
\begin{equation}
\label{f4}
\Psi(\boldsymbol{\rho},\boldsymbol{\lambda},\boldsymbol{\sigma}) = \mathcal{A}\sum_{I=1}^K C_I \exp\Bigl[-\frac{1}{2}
\bigl(A_{11}(I)\boldsymbol{\rho}^2 + 2A_{12}(I)\boldsymbol{\rho}\boldsymbol{\lambda} + A_{22}(I)\boldsymbol{\lambda}^2+
\end{equation}
\begin{displaymath}
2A_{13}(I)\boldsymbol{\rho}\boldsymbol{\sigma} + 
2A_{23}(I)\boldsymbol{\lambda}\boldsymbol{\sigma} + A_{33}(I)\boldsymbol{\sigma}^2\bigr)\Bigr],
\end{displaymath}
where $A_{ij}(I)$ is the matrix of nonlinear variational parameters, $C_I$ are linear variational parameters. 
The operator $\mathcal{A}$ denotes the appropriate symmetrization or antisymmetrization of the wave function, i.e., 
an operator that creates a final wave function with the correct permutation symmetry. In many systems considered 
in this study, there are pairs of identical fermions. The total wave function of the system of particles, which includes 
the coordinate and spin wave functions, must be antisymmetric under the permutation of identical fermions.

When determining the energies of lepton bound states by the variational method, it is necessary to find such values 
of the expansion parameters and coefficients for which the average value of the Hamiltonian is minimal. 
To obtain the bound state energies, the Schr\"odinger equation with Coulomb interaction of four particles is reduced 
to solving a matrix eigenvalue problem of the form:
\begin{equation}
HC = E^\lambda BC,
\label{f5}
\end{equation}
where the matrix elements of the Hamiltonian $H_{ij} = \langle\psi_i|H|\psi_j\rangle$ and the normalization coefficients 
$B_{ij} = \langle\psi_i|\psi_j\rangle$ are calculated analytically using variational wave functions \eqref{f4}, and $E^\lambda$ 
is one of the energy eigenvalues. The upper bound for the energy of the state of the four-particle system in variational 
approach is determined by the smallest eigenvalue of the generalized eigenvalue problem \eqref{f5}.

The normalization condition for the wave function \eqref{f4} contains an integral that is calculated analytically and has the form:
\begin{equation}
B = \int d\boldsymbol{\rho} \int d\boldsymbol{\lambda} \int d\boldsymbol{\sigma} \exp\left[-\frac{1}{2}(B_{11}\boldsymbol{\rho}^2 + 2B_{12}\boldsymbol{\rho}\boldsymbol{\lambda} + B_{22}\boldsymbol{\lambda}^2 + 2B_{13}\boldsymbol{\rho}\boldsymbol{\sigma} + 2B_{23}\boldsymbol{\lambda}\boldsymbol{\sigma} + B_{33}\boldsymbol{\sigma}^2)\right] = 
\end{equation}
\begin{displaymath}
=\frac{(2\pi)^{9/2}}{(\det B)^{3/2}},
\label{f6}
\end{displaymath}
where the parameter matrix $B_{ij}$ is introduced, with the determinant
\begin{equation}
\det B = B_{11}B_{22}B_{33} - B_{11} B_{23}^2 - B_{12}^2B_{33} + 2 B_{12}B_{13} B_{23} - B_{13}^2 B_{22}, \quad B_{kl}(I,J) = A_{kl}(I) + A_{kl}(J).
\label{f7}
\end{equation}

As a result, the normalization coefficient of the wave function \eqref{f4} in the coordinate representation is determined 
by the formula:
\begin{equation}
\mathcal{N} = \sum_{I=1}^K\sum_{J=1}^K C_I C_J \frac{(2\pi)^{9/2}}{(\det B(I,J))^{3/2}},
\label{f8}
\end{equation}
where the matrix elements $B_{kl}$ in \eqref{f8} depend on two summation indices.

Let us also calculate the matrix elements of kinetic and potential energy, necessary to solve 
the matrix eigenvalue problem. 
The kinetic energy operator consists of three terms, defined by the Laplace operators for variables $\boldsymbol{\rho}$, $\boldsymbol{\lambda}$, $\boldsymbol{\sigma}$. All these matrix elements are expressed in terms of variational parameters 
and have the following form:
\begin{equation}
\langle\Delta_\rho\rangle = \frac{48\sqrt{2}\pi^{9/2}}{\det B^{5/2}} \Bigl[A_{11}^2(B_{22}B_{33}-B_{23}^2)+A_{12}^2(B_{11}B_{33}-B_{13}^2)+A_{13}^2(B_{11}B_{22}-B_{12}^2)-A_{11}\det B
\end{equation}
\begin{displaymath}
+2A_{11}A_{12}(B_{13}B_{23}-B_{12}B_{33})+2A_{11}A_{13}(B_{12}B_{23}-B_{13}B_{22})+2A_{12}A_{13}(B_{12}B_{13}-B_{11}B_{23})\Bigr], 
\label{f9}
\end{displaymath}
\begin{equation}
\label{f10}
\langle\Delta_\lambda\rangle = \frac{48\sqrt{2}\pi^{9/2}}{\det B^{5/2}} \Bigl[A_{12}^2(B_{22}B_{33}-B_{23}^2)+A_{22}^2(B_{11}B_{33}-B_{13}^2)+A_{23}^2(B_{11}B_{22}-B_{12}^2)-A_{22}\det B
\end{equation}
\begin{displaymath}
+ 2A_{22}A_{12}(B_{13}B_{23}-B_{12}B_{33})+2A_{22}A_{23}(B_{12}B_{13}-B_{11}B_{23})+2A_{12}A_{23}(B_{12}B_{23}-B_{13}B_{22})\Bigr],
\end{displaymath}
\begin{equation}
 \label{f11}
\langle\Delta_\sigma\rangle = \frac{48\sqrt{2}\pi^{9/2}}{\det B^{5/2}} \Bigl[A_{13}^2(B_{22}B_{33}-B_{23}^2)+A_{23}^2(B_{11}B_{33}-B_{13}^2)+A_{33}^2(B_{11}B_{22}-B_{12}^2)-A_{33}\det B
\end{equation}
\begin{displaymath}
+ 2A_{13}A_{23}(B_{13}B_{23}-B_{12}B_{33})+2A_{33}A_{13}(B_{12}B_{23}-B_{13}B_{22})+2A_{33}A_{23}(B_{12}B_{13}-B_{11}B_{23})\Bigr].
\end{displaymath}

The matrix elements of potential energy in the nonrelativistic approximation are the matrix elements of pairwise Coulomb interactions. 
The direct calculation of the corresponding integrals with wave functions \eqref{f4} gives the following results:
\begin{equation}
\left\langle\frac{1}{|\mathbf{r}_3-\mathbf{r}_1|}\right\rangle= \frac{32\pi^4}{\det B\sqrt{B_{11}B_{33}-B_{13}^2-\frac{m_2}{m_{12}}(2B_{12}B_{33}-2B_{13}B_{23}-\frac{m_2}{m_{12}}(B_{22}B_{33}-B_{23}^2))}}, \label{f12}
\end{equation}
\begin{equation}
\left\langle\frac{1}{|\mathbf{r}_1-\mathbf{r}_2|}\right\rangle= \frac{32\pi^4}{\det B\sqrt{(B_{22}B_{33}-B_{23}^2)}}, \label{f13} 
\end{equation}
\begin{equation}
\left\langle\frac{1}{|\mathbf{r}_3-\mathbf{r}_4|}\right\rangle= \frac{32\pi^4}{\det B\sqrt{B_{11}B_{22}-B_{12}^2-\frac{m_{12}}{m_{123}}(2B_{12}B_{13}-2B_{11}B_{23}-\frac{m_{12}}{m_{123}}(B_{11}B_{33}-B_{13}^2))}}, \label{f14}
\end{equation}
\begin{equation}
\left\langle\frac{1}{|\mathbf{r}_3-\mathbf{r}_2|}\right\rangle= \frac{32\pi^4}{\det B\sqrt{B_{11}B_{33}-B_{13}^2+\frac{m_{1}}{m_{12}}(2B_{12}B_{33}-2B_{13}B_{23}+\frac{m_{1}}{m_{12}}(B_{22}B_{33}-B_{23}^2))}}, \label{f15}
\end{equation}
\begin{equation}
\label{f16}
\left\langle\frac{1}{|\mathbf{r}_1-\mathbf{r}_4|}\right\rangle= \frac{32\pi^4}{\det B\sqrt{F_{14}}}, \quad F_{14} = \frac{m_3}{m_{123}} (2 B_{12} B_{13}-2 B_{11} B_{23})+\frac{m_3^2}{m_{123}^2} (B_{11} B_{33}-B_{13}^2)
\end{equation}
\begin{displaymath}
+ B_{11} B_{22}-B_{12}^2+\frac{m_2}{m_{12}} \left(\frac{m_3}{m_{123}} (2 B_{13} B_{23}-2 B_{12} B_{33})+2 B_{12} B_{23}-2 B_{13} B_{22}\right)+\frac{m_2^2}{m_{12}^2} (B_{22} B_{33}-B_{23}^2),
\end{displaymath}
\begin{equation}
 \label{f17}
\left\langle\frac{1}{|\mathbf{r}_2-\mathbf{r}_4|}\right\rangle= \frac{32\pi^4}{\det B\sqrt{F_{24}}}, \quad F_{24} = \frac{m_3}{m_{123}} (2 B_{12} B_{13}-2 B_{11} B_{23})+\frac{m_3^2}{m_{123}^2} (B_{11} B_{33}-B_{13}^2)
\end{equation}
\begin{displaymath}
+ B_{11} B_{22}-B_{12}^2-\frac{m_1}{m_{12}} \left(\frac{m_3}{m_{123}} (2 B_{13} B_{23}-2 B_{12} B_{33})+2 B_{12} B_{23}-2 B_{13} B_{22}\right)+\frac{m_1^2}{m_{12}^2} (B_{22} B_{33}-B_{23}^2).
\end{displaymath}

Let's also consider calculating the hyperfine structure of the energy spectrum of four-particle systems.
Changing the particle composition leads to changes in the hyperfine structure itself, which, in the presence of identical fermions, acquires constraints due to the structure of the spin wave function. As an example, we will consider the hyperfine structure of the positronium hydride ($HPs$) system, which has two identical electrons.
The general Hamiltonian of the hyperfine interaction is the following:
\begin{equation}
\Delta V^{\text{hfs}}(\boldsymbol{\rho},\boldsymbol{\lambda},\boldsymbol{\sigma}) = a_{12}(\mathbf{s}_1\mathbf{s}_2) + a_{34}(\mathbf{s}_3\mathbf{s}_4) + a_{13}(\mathbf{s}_1\mathbf{s}_3) + a_{14}(\mathbf{s}_1\mathbf{s}_4) + a_{23}(\mathbf{s}_2\mathbf{s}_3) + 
a_{24}(\mathbf{s}_2\mathbf{s}_4),
\label{f18}
\end{equation}
\begin{equation}
a_{12} = -\frac{8\pi\alpha}{3m_e m_p}\delta(\mathbf{r}_{12}), \quad a_{13} = \frac{8\pi\alpha}{3m_e m_p}
\delta(\mathbf{r}_{13}), \quad a_{14} = -\frac{8\pi\alpha}{3m_e m_p}\delta(\mathbf{r}_{14}),
\label{f25}
\end{equation}
\begin{equation}
a_{23} = -\frac{8\pi\alpha}{3m_e^2}\delta(\mathbf{r}_{23}), \quad a_{24} = \frac{8\pi\alpha}{3m_e^2}
\delta(\mathbf{r}_{24}), \quad a_{34} = -\frac{8\pi\alpha}{3m_e^2}\delta(\mathbf{r}_{34}),
\end{equation}
where $\mathbf{s}_i$ $(i=1, 2, 3, 4)$ are the spin operators of the particles. The particle numbering corresponds to their order 
in the HPs molecule (proton-electron-positron-electron).

When constructing spin wave functions for various HPs states, we will perform the addition of particle spins 
in the following sequence: first, we add the spins of the first and second electrons $\mathbf{S}_{12}=\mathbf{s}_1+\mathbf{s}_2$, 
then the spins of the proton and positron $\mathbf{S}_{34}=\mathbf{s}_3+\mathbf{s}_4$ and finally we obtain the total spin 
of the system as $\mathbf{S}=\mathbf{S}_{12}+\mathbf{S}_{34}$. If the spin of the electron pair $\mathbf{S}_{12}=0$, 
then the antisymmetric spin wave function of the system $\chi^{S_{12}S_{34}}_{SS_z}$ under electron permutation has the form:
\begin{equation}
\chi^{00}_{00} = \frac{1}{2}\left(\uparrow\downarrow\uparrow\downarrow+\downarrow\uparrow\downarrow\uparrow - \uparrow\downarrow\downarrow\uparrow-\downarrow\uparrow\uparrow\downarrow\right), \quad \chi^{01}_{11} = \frac{1}{\sqrt{2}}\left(\uparrow\downarrow\uparrow\uparrow-\downarrow\uparrow\uparrow\uparrow\right), \quad 
\uparrow=\binom{1}{0}, \quad \downarrow=\binom{0}{1}.
\label{f26}
\end{equation}
The functions \eqref{f26} correspond to two possible values of the spin of the proton-positron pair $S_{34}=0, 1$. 
Other spin states of the particles are described by the following wave functions:
\begin{align}
\chi_{00}^{11} &= \frac{1}{\sqrt{12}}\left(2\uparrow\uparrow\downarrow\downarrow-\uparrow \downarrow   \uparrow \downarrow - 
\uparrow  \downarrow\downarrow \uparrow-\downarrow\uparrow\uparrow\downarrow-\downarrow \uparrow \downarrow\uparrow+2 
\downarrow\downarrow \uparrow\uparrow\right), \label{f27} \\
\chi_{11}^{11} &= \frac{1}{2}\left(\uparrow\uparrow\uparrow\downarrow+ \uparrow\uparrow  \downarrow\uparrow-\uparrow 
\downarrow \uparrow\uparrow-\downarrow\uparrow\uparrow\uparrow\right), \label{f28} \\
\chi_{22}^{11} &= \uparrow\uparrow\uparrow\uparrow, \quad \chi^{10}_{11} = \frac{1}{\sqrt{2}}
\left(\uparrow\uparrow\uparrow\downarrow-\uparrow\uparrow\downarrow\uparrow\right). \label{f29}
\end{align}
The functions \eqref{f27}-\eqref{f29} correspond to the spin value of the pair of two-electrons $S_{12}=1$.

We transform the matrix element of the Hamiltonian \eqref{f18}, taking into account that 
$\langle\delta(\mathbf{r}_{12})\rangle = \langle\delta(\mathbf{r}_{14})\rangle$, $\langle\delta(\mathbf{r}_{23})\rangle = \langle\delta(\mathbf{r}_{34})\rangle$:
\begin{equation}
\Delta E^{\text{hfs}} = -\frac{8\pi\alpha}{3m_e m_p}(\mathbf{s}_1\mathbf{s}_{24})\langle\delta(\mathbf{r}_{12})\rangle - 
\frac{8\pi\alpha}{3m_e^2}(\mathbf{s}_3\mathbf{s}_{24})\langle\delta(\mathbf{r}_{23})\rangle + \frac{8\pi\alpha}{3m_e m_p}(\mathbf{s}_1\mathbf{s}_{3})\langle\delta(\mathbf{r}_{13})\rangle.
\label{f30}
\end{equation}

If the total spin of the electron pair $\mathbf{s}_{24}=0$, then the hyperfine structure of the spectrum 
is determined by the interaction of the spins of the proton and positron:
\begin{equation}
\Delta E^{\text{hfs}}=\frac{8\pi\alpha^4 m_e^2}{3m_p}(\mathbf{s}_1\mathbf{s}_{3})
\langle\delta(\mathbf{r}_{13})\rangle.
\label{f31}
\end{equation}

The matrix element of the Dirac $\delta$ function in \eqref{f31} has the following form:
\begin{equation}
\label{f32}
\langle\delta(\mathbf{r}_{13})\rangle = 8\pi^3\Bigl\{\frac{m_3^2}{m_{123}^2}(B_{11}B_{33}-B_{13}^2)+
\frac{m_1^2}{m_{12}^2}(B_{22}B_{33}-B_{23}^2)+B_{11}B_{22}-B_{12}^2 + 
\end{equation}
\begin{displaymath}
2\frac{m_3}{m_{123}}(B_{12}B_{13}-B_{11}B_{23}) + 2\frac{m_1}{m_{12}}(B_{13}B_{22}-B_{12}B_{23})+
2\frac{m_1}{m_{12}}\frac{m_3}{m_{123}}(B_{12}B_{33}-B_{13}B_{23})\Bigr\}^{-3/2}.
\end{displaymath}

The numerical value of this matrix element is $\langle\delta(\mathbf{r}_{13})\rangle = 0.001582$, 
which gives the hyperfine 
splitting value $\Delta E^{\text{hfs}}(\text{HPs}) = 2.692~\text{MHz}$. Similarly, for the MuPs system, 
where the bound state 
arises when the total spin of the two electrons is zero, the hyperfine splitting is 
$\Delta E^{\text{hfs}}(\text{MuPs}) = 22.458~\text{MHz}$ for $\langle\delta(\mathbf{r}_{13})\rangle = 0.001684$.

To study the internal structure of four-particle bound states, we calculate the mean square distances 
between particles.
The corresponding matrix elements are calculated analytically and the calculation results are presented 
in Appendix~\ref{app1}.
As an example, we present the values of the mean square distances between particles in the 
muonium-positronium (MuPs) molecule, numbering the particles as follows: $e^-\to 1$, $\mu^+\to 2$,
$e^-\to 3$, $e^+\to 4$.
Then, using relations \eqref{a1}-\eqref{a6}, we obtain:
\begin{equation}
\label{f33}
\delta_{12}=\sqrt{\langle({\bf r}_1-{\bf r}_2)^2\rangle}=2.80~(1.48\times 10^{-8}~cm),~
\delta_{13}=\sqrt{\langle({\bf r}_1-{\bf r}_3)^2\rangle}=4.00~(2.12\times 10^{-8}~cm),~
\end{equation}
\begin{equation}
\label{f34}
\delta_{14}=\sqrt{\langle({\bf r}_1-{\bf r}_4)^2\rangle}=3.95~(2.09\times 10^{-8}~cm),~
\delta_{24}=\sqrt{\langle({\bf r}_2-{\bf r}_4)^2\rangle}=3.95~(2.09\times 10^{-8}~cm).
\end{equation}

The average distances between particles are presented in \eqref{f33}-\eqref{f34} in electron atomic units and cm 
(in brackets).

\section{Conclusion}

In this work, studies of the energy levels of three-particle and four-particle lepton bound states within the stochastic 
variational method in quantum electrodynamics are carried out. The variational functions are chosen in the Gaussian form. 
The Schr\"odinger equation for the bound state of particles is reduced to a matrix equation, which is solved numerically. 
For the numerical solution of the matrix equation in the case of four particles, a program is written in the MATLAB system. 
The original Varga-Suzuki program in Fortran from \cite{p7} and our program for calculating three-particle energy levels 
of previous works \cite{p8,p9,p10} are taken as the basis for writing the program. In general, the efficiency 
of the variational method in solving the equation \eqref{f5} depends on the algorithms used to optimize the nonlinear 
parameters in the wave function. In this work, we have extended the previously used variational method \cite{p8,p9,p10} 
to the case of the Coulomb systems of four-particles. For this, the Jacobi coordinates and the bound state wave function 
\eqref{f4} are used.
The use of trial Gaussian functions in the variational method allows us to obtain values of the energies of bound states 
of leptons that are in good agreement with previous calculations performed in other approaches.

To calculate the energy of a specific state, we run the program 10 times with a minimum basis of 200 functions,
varying the boundaries to generate nonlinear variational parameters. Having chosen the boundaries so that the minimum 
energy is achieved at this stage, we run several programs with the same parameters, increasing the basis
in the decomposition \eqref{f4} to approximately 800 functions. At each step of the program, the basis is incremented 
by one, and refinement cycles are performed. The number of such cycles (usually 1000) is specified in the input file 
at the beginning of the program.

The results of the numerical calculations and their comparison with some results of other authors are presented 
in Table~\ref{tb1} in electronic atomic units (e.a.u.): $1~\text{e.a.u.} = 27.211385~\text{eV}$. Table~\ref{tb1} 
includes results for the four-particle systems HPs (positronium hydride) and HMu (muonium hydride). 
The obtained results include the total energies of bound states and the matrix elements of the system's Hamiltonian. 
In our calculations, we use the following particle masses:
$m_p/m_e=1836.152 673 426(32)$, $m_\mu/m_e= 206.768 2827(46)$ \cite{codata}.

In general, our results agree with calculations by other authors \cite{p11,p11a,p12,p13,p14,p14a}. Some difference 
in results (no more than $0.04\%$) is due, in our opinion, to two circumstances. First, different basis sizes were used. 
Second, the formulation of the variational approach itself (choice of coordinates, type of basis functions) is somewhat 
different. For example, in the works \cite{p13,p14,p14a}, perimetric relative coordinates and exponential variational 
functions were chosen.

\begin{table}[htbp]
\centering
\caption{Binding energies of lepton states in electronic atomic units (e.a.u.).}
\vspace{3mm}
\label{tb1}
\begin{tabular}{|l|c|c|c|c|c|c|c|c|}
\toprule
Bound & This work & \cite{p5} & \cite{p5a} & \cite{p11} & \cite{p14} & \cite{p19} & \cite{p20} & \cite{p21} \\
 state &   &   &   &   &   &   &  &   \\  \hline
%\midrule
$\text{Mu}_2$ & $-1.140230$ & -- & -- & $-1.141013$ & -- & -- & -- & -- \\ \hline
$\text{Ps}_2$ & $-0.515982$ & $-0.516003$ & -- & -- & -- & $-0.515989$ & -- & -- \\ \hline
$\text{MuPs}$ & $-0.786262$ & -- & -- & -- & $-0.786317$ & -- & -- & -- \\ \hline
$\text{Ps}^-$ & $-0.261954$ & -- & -- & -- & -- & -- & $-0.262005$ & -- \\ \hline
$\text{Mu}^-$ & $-0.525054$ & -- & -- & -- & $-0.525054$ & -- & -- & -- \\ \hline
$\text{HPs}$ & $-0.788819$ & $-0.788867$ & $-0.788870$ & -- & -- & -- & -- & -- \\ \hline
$\text{HMu}$ & $-1.148355$ & -- & -- & -- & -- & -- & -- & $-1.150187$ \\ \hline
$\text{TMu}_2$ & $-106.6603$ & -- & -- & -- & -- & -- & -- & -- \\ \hline
%\bottomrule
\end{tabular}
\end{table}

The results of the performed study show that the energy levels of bound states of various leptons (primarily 
with $e^-$, $e^+$, $\mu^+$, $\mu^-$) are determined with good numerical accuracy
(In Table ~\ref{tb1} they are presented with an accuracy of 6 decimal places.) These results can be improved by taking 
into account various corrections to the nonrelativistic results, primarily relativistic corrections of order of $O(\alpha^2)$. 
They will contribute to the last two decimal places. Such lepton bound states can be observed experimentally. For example, 
bound states of electrons and positrons have been discovered (three-particle state -- positronium ion $\text{Ps}^-$, 
four-particle state -- positronium molecule $\text{Ps}_2$) \cite{p4,p5,p15,p16,p18}. 
The internal structure of the four-particle systems under study can be represented as two-particle clusters formed by two neutral 
atoms interacting weakly with each other. This interaction between these two neutral atoms is described by van der Waals attractive 
forces, which, at certain particle masses, are sufficient to form a bound state of the entire four-particle system.

Experimental studies of such systems include not only determining energy levels but also measuring decay widths. Positronium hydride 
HPs was discovered in \cite{HPs} while studying collisions of positrons and methane molecules. The next step could be conducting 
an experiment to create a real muonium-positronium system MuPs, observing its decays and measuring some of its properties. 
One of the main problems in creating condensed muonium matter is the very short lifetime of the muon \cite{p23}. 
The transition from positronium and muonium to three-particle and four-particle bound states of leptons when studying energy 
levels is quite natural, as it opens up new possibilities for testing the theory of bound states in the Standard Model. 
Table~\ref{tb1} includes the result for the molecule $\text{TMu}_2$ (TMu is true muonium $\mu^+\mu^-$), consisting 
of $(\mu^+\mu^-\mu^+\mu^-)$.
Further studies of lepton bound states may aim to refine the obtained results by taking into account radiative 
and relativistic corrections.

\begin{acknowledgments}
The authors are grateful to V. I. Korobov for useful discussions.
This work was supported by the Russian Science Foundation (Grant No. 25-72-00029) (F.A.M.).
\end{acknowledgments}

\appendix
\section{Root mean square distances between particles.}
\label{app1} 

To calculate the mean square distances between particles, we use the explicit form of the wave function 
of the system \eqref{f4}. The results of analytical integration are as follows:
\begin{equation}
\label{a1}
\langle({\bf r}_1-{\bf r}_2)^2\rangle=\frac{48\sqrt{2}\pi^{\frac{9}{2}}(B_{22}B_{33}-B_{23}^2)}{\det B^{\frac{5}{2}}},
\end{equation}
\begin{equation}
\label{a2}
\langle({\bf r}_3-{\bf r}_1)^2\rangle=\frac{48\sqrt{2}\pi^{\frac{9}{2}}}{\det B^{\frac{5}{2}}}
[B_{11}B_{33}-B_{13}^2+2\frac{m_2}{m_{12}}(B_{13}B_{23}-B_{12}B_{33})+\frac{m_2^2}{m_{12}^2}(B_{22}B_{33}-B_{23}^2)],
\end{equation}
\begin{equation}
\label{a3}
\langle({\bf r}_3-{\bf r}_4)^2\rangle=\frac{48\sqrt{2}\pi^{\frac{9}{2}}}{\det B^{\frac{5}{2}}}
[B_{11}B_{22}-B_{12}^2+2\frac{m_{12}}{m_{123}}(B_{11}B_{23}-B_{12}B_{13})+\frac{m_{12}^2}{m_{123}^2}
(B_{11}B_{33}-B_{13}^2)],
\end{equation}
\begin{equation}
\label{a4}
\langle({\bf r}_2-{\bf r}_3)^2\rangle=\frac{48\sqrt{2}\pi^{\frac{9}{2}}}{\det B^{\frac{5}{2}}}
[B_{11}B_{33}-B_{13}^2+2\frac{m_{1}}{m_{12}}(B_{12}B_{33}-B_{13}B_{23})+\frac{m_{1}^2}{m_{12}^2}
(B_{22}B_{33}-B_{23}^2)],
\end{equation}
\begin{equation}
\label{a5}
\langle({\bf r}_1-{\bf r}_4)^2\rangle=\frac{48\sqrt{2}\pi^{\frac{9}{2}}}{\det B^{\frac{5}{2}}}
[B_{11}B_{22}-B_{12}^2+2\frac{m_{2}}{m_{12}}(B_{12}B_{23}-B_{13}B_{22})+\frac{m_{2}^2}{m_{12}^2}
(B_{22}B_{33}-B_{23}^2)+
\end{equation}
\begin{displaymath}
2\frac{m_{2}}{m_{12}}\frac{m_3}{m_{123}}(B_{13}B_{23}-B_{12}B_{33})+
2\frac{m_{3}}{m_{123}}(B_{12}B_{13}-B_{11}B_{23})+\frac{m_{3}^2}{m_{123}^2}
(B_{11}B_{33}-B_{13}^2)]
\end{displaymath}
\begin{equation}
\label{a6}
\langle({\bf r}_2-{\bf r}_4)^2\rangle=\frac{48\sqrt{2}\pi^{\frac{9}{2}}}{\det B^{\frac{5}{2}}}
[B_{11}B_{22}-B_{12}^2+2\frac{m_{3}}{m_{123}}(B_{12}B_{13}-B_{11}B_{23})+\frac{m_{3}^2}{m_{123}^2}
(B_{11}B_{33}-B_{13}^2)+
\end{equation}
\begin{displaymath}
2\frac{m_{1}}{m_{12}}\frac{m_3}{m_{123}}(B_{12}B_{33}-B_{13}B_{23})+
2\frac{m_{1}}{m_{12}}(B_{13}B_{22}-B_{12}B_{23})+\frac{m_{1}^2}{m_{12}^2}
(B_{22}B_{33}-B_{23}^2)].
\end{displaymath}

Using \eqref{a1}-\eqref{a6} one can find the numerical values of the average distances between particles.

\end{document}